\documentclass{optica-article}

\journal{opticajournal} 

\articletype{Research Article}

\usepackage{lineno}
\usepackage{float}
\newcommand{\dd}{\mathrm{d}}

\begin{document}

\title{Comparative analysis of optical rectification-based multicycle terahertz pulse generation techniques}

\author{Luis Nasi,\authormark{1,2} Gergő Illés,\authormark{1} János Hebling,\authormark{1,2,3} and György Tóth\authormark{1,2,*}}

\address{\authormark{1}Institute of Physics, University of Pécs, Pécs, Hungary\\
\authormark{2}Szentágothai Research Center, University of Pécs, Pécs, Hungary\\
\authormark{3}HUN-REN–PTE High Field Terahertz Research Group, Pécs, Hungary}

\email{\authormark{*}tothgy@fizika.ttk.pte.hu} 


\begin{abstract*} 
Numerical investigations of multicycle terahertz pulse generation based on wafer stack and tilted pulse front pumped lithium niobate setups concerning the conversion efficiency and the resulting temporal shape and spectrum were performed. Pumping by Fourier-limited-, chirped-, and intensity-modulated pulses, as well as pulse sequences was considered. Wafer stacks give a maximum efficiency when the pump pulse duration is 0.243 times the temporal period of the generated THz pulse. In wafer stacks, increasing the number of pulses in a sequence drastically increases the efficiency. A five-period wafer stack tailored for a 0.15 THz frequency achieves a maximum efficiency of almost 0.1\% when pumped by a single, 1.5 ps pulse with an intensity of 100 GW$^2$/cm$^2$, and 0.3\% when pumped by a sequence of five pulses. In a similar wafer stack of 10 periods, the efficiency with the same pulse sequence reaches 0.7\%. For a large pump pulse number, the efficiency increase approaches a factor equal to the number of wafer pairs. For a given number of THz cycles, the highest efficiency is obtained when the number of wafer stack periods and pump pulses are equal to half the number of cycles. For relatively small cycle numbers, tilted-pump-front setups were found to yield the highest conversion efficiencies. 
\end{abstract*}

\section{Introduction}
In the last two decades, significant progress has been made in the generation and application of high energy terahertz (THz) pulses \cite{Salen2019}, mostly owing to the progress in techniques based on velocity-matched optical rectification – such as tilted pulse front pumping (TPFP) \cite{Hebling2002} of lithium niobate (LN) or semiconductors \cite{Mbithi2022,Toth2023,Fueloep2016} – and development of high quality organic nonlinear optical crystals \cite{Jazbinsek2019}. These THz sources usually generate near-single-cycle pulses. However, some promising THz pulse driven electron acceleration schemes require narrowband, multicycle THz pulses with high electric field strength \cite{Matlis2018}. In most multicycle THz pulse generation experiments, periodically poled lithium niobate (PPLN) crystal \cite{Carbajo2015,Ahr2017,Jolly2019,Olgun2022}, or a wafer stack structure (WS) \cite{Lemery2020,Mosley2023} made of periodically inverted wafers of lithium niobate has been used. While some of these techniques can also be used to generate up to hundreds of cycles, in our paper we aim to investigate and compare techniques useful for a lower number of cycles, in the 2-20 range.

In the simplest case, a single laser pulse with an approximately Gaussian temporal shape is used for pumping the LN or semiconductor nonlinear optical material \cite{Carbajo2015,Lemery2020,Mosley2023,Vodopyanov2006 }. The generated $\nu_0$ THz frequency is given by the

\begin{equation}
\nu_0 =
 \frac{c}{\Lambda \left(n_{\mathit{THz}}^{ph}(\nu_0)-n_{p}^{gr}\right)}
\label{eq1}
\end{equation}
equation, where $c$ is the speed of light, $\Lambda$ is the period length of the wafer stack (twice the thickness of one wafer), $ n_{\mathit{THz}}^{ph}$ is the THz phase refractive index and $n_p^{gr}$ is the optical group refractive index.

However, higher conversion efficiency can be achieved using a sequence of pump pulses \cite{Ravi2016,Stummer2020,Matlis2023} or intensity-modulated pump pulses \cite{Ahr2017,Jolly2019,Olgun2022} instead of a single pump pulse. The most widely used technique for the generation of intensity-modulated pump pulses is the chirp-and-delay (C\&D) technique, in which case the chirped pulse is split into two pulses by a beamsplitter, and then the intensity-modulated pulse is created by the superposition of these two pulses with a relative time delay \cite{Ahr2017,Jolly2019}. The amount of the chirp and the delay together determine the modulation frequency of the pulse according to the following equation \cite{Toth2017}:

\begin{equation}
\Delta \nu =
 \frac{4\ln 2}{\tau^2 \tau_0}\sqrt{\tau^2 -\tau_0^2}~ \Delta \tau,
\label{eq2}
\end{equation}
where $\tau_0$ and $\tau$ are the Fourier-transform-limited (FL) and chirped pulse durations, respectively, and $\Delta \tau$ is the time delay. This kind of chirped and intensity-modulated pulses can also be generated directly by the superposition of the chirped signal and idler beams of an optical parametric chirped pulse amplifier \cite{Toth2017}.

The other possibility for producing intensity-modulated pulses is superposing two beams having narrowband spectra (TNB) with different mean frequencies \cite{Olgun2022,Nugraha2018}. In this case, the modulation frequency is equal to the difference of the two mean frequencies.

THz generation by intensity-modulated pump pulses in a TPFP scheme has been demonstrated \cite{Chen2011}, but their use is not widespread yet. The waveform of the generated THz pulse in phase-matched optical rectification is that of the time derivative of the intensity of the pump pulse \cite{Wynne2005}. In this way, multicycle THz pulses can be generated by using intensity-modulated pulse or pulse sequence for the pumping of TPFP scheme.

\section{The investigated techniquies}
In this work, both the TPFP and the WS THz generation setups have been investigated when used with different type of optical pump pulses. These pulses are schematically illustrated in Fig. 1. 

\begin{figure}[H]
\centering\includegraphics[width=13cm]{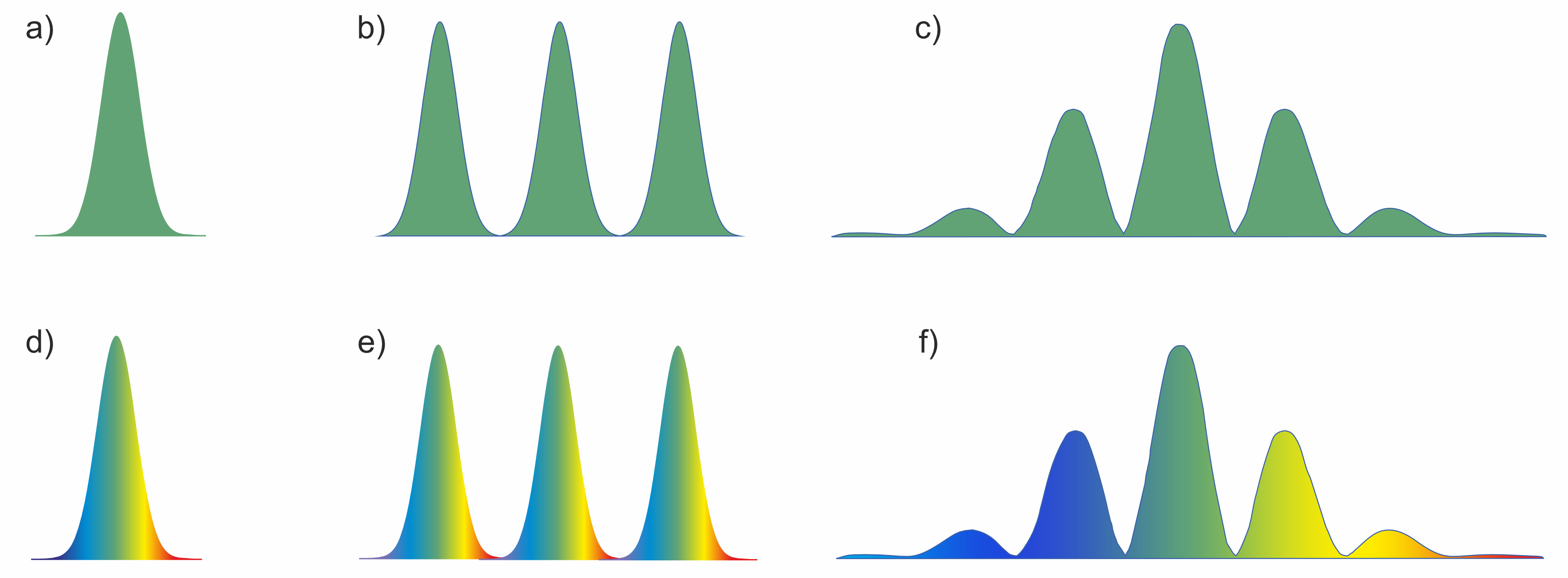}
\caption{Schematic illustration of the considered optical pump pulses: (a) single FL Gaussian pulse, (b) FL Gaussian pulse sequence, (c) intensity-modulated pulse created with the TNB technique, (d) single chirped Gaussian pulse, (e) chirped Gaussian pulse sequence, (f) intensity-modulated pulse created with the C\&D technique.}
\end{figure}

For the cases of FL Gaussian pulse (Fig. 1a), FL pulse sequence (Fig. 1b), TNB intensity-modulated Gaussian pulse (Fig. 1c), chirped Gaussian pulse (Fig 1d), chirped pulse sequence (Fig. 1e) and C\&D intensity-modulated pulse (Fig. 1f), respectively, the temporal shape of the pump intensity is given as:

\begin{equation}
I(t) =
 I_0 e^{-4\ln (2)\frac{t^2}{\tau^2_0}}.
\end{equation}

\begin{equation}
I(t) =
 I_0 \sum_{j=1}^N e^{-4\ln (2)\frac{[t+(j-N/2)\Delta \tau ]^2}{\tau^2_0}}.
\end{equation}

\begin{equation}
I(t) =
 \frac{1}{2} I_0 e^{-4\ln (2)\frac{t^2}{\tau^2_0}}\left[ e^{i(\omega_0+\pi \Delta \nu )t}+e^{i(\omega_0-\pi \Delta \nu )t}\right]^2.
\end{equation}

\begin{equation}
I(t) =
 I_0 \frac{\tau}{\tau_0} \mathcal{F}^{-1}\left\{ \mathcal{F}\left\{e^{-2\ln (2)\frac{t^2}{\tau^2_0}}e^{i\omega_0 t}\right\} e^{i\frac{\mathit{GDD}}{2}(\omega -\omega_0)^2}\right\}.
\end{equation}

\begin{equation}
I(t) =
 I_0\frac{\tau}{\tau_0} \sum_{j=1}^N \mathcal{F}^{-1}\left\{ \mathcal{F}\left\{e^{-2\ln (2)\frac{t^2}{\tau^2_0}}e^{i\omega_0 t}\right\} e^{i\frac{\mathit{GDD}}{2}(\omega -\omega_0)^2} e^{(j-N/2)\frac{GD}{2}(\omega -\omega_0)}\right\}.
\end{equation}

\begin{equation}
I(t) =
 \frac{1}{2}I_0\frac{\tau}{\tau_0} \mathcal{F}^{-1}\left\{ \mathcal{F}\left\{e^{-2\ln (2)\frac{t^2}{\tau^2_0}}e^{i\omega_0 t}\right\} e^{i\frac{\mathit{GDD}}{2}(\omega -\omega_0)^2} \left[ e^{i\frac{\Delta \tau }{2}(\omega -\omega_0)}+e^{-i\frac{\Delta \tau }{2}(\omega -\omega_0)}\right] \right\},
\end{equation}
where
\begin{equation}
\mathit{\mathit{GDD}} =
 \frac{\tau_0^2}{4\ln (2)\sqrt{\frac{\tau^2}{\tau_0^2}-1}}
\end{equation}
is the group delay dispersion. In the C\&D case (Eq. 8) the modulation frequency is the following \cite{Toth2017}:
\begin{equation}
\Delta \nu =
 \frac{1}{\pi}\left( 2\mathit{GDD}+\frac{\tau_0^4}{8\ln^2 (2) ~ \mathit{GDD}} \right)^{-1}\Delta \tau .
\end{equation}
In this article, by pulse sequence we will refer to an equidistant series of identical pulses (cases in Fig. 1 b, e).

In Fig. 2, pump pulses generated by the C\&D and TNB techniques are presented. In the calculations, the used parameters were $I_0 = 100~\mathrm{GW/cm^2,~} \tau_0 = 500~\mathrm{fs,~} \tau = 33~\mathrm{ps~and~} \Delta \nu = 0.15~\mathrm{THz.}$  Although the temporal shapes of the intensities (Fig. 2a) are practically the same in the two cases, the spectrums (Fig. 2b) differ drastically. For visibility, the inset of Fig. 2b offers a magnified view of the spectrums.

\begin{figure}[H]
\centering\includegraphics[width=13cm]{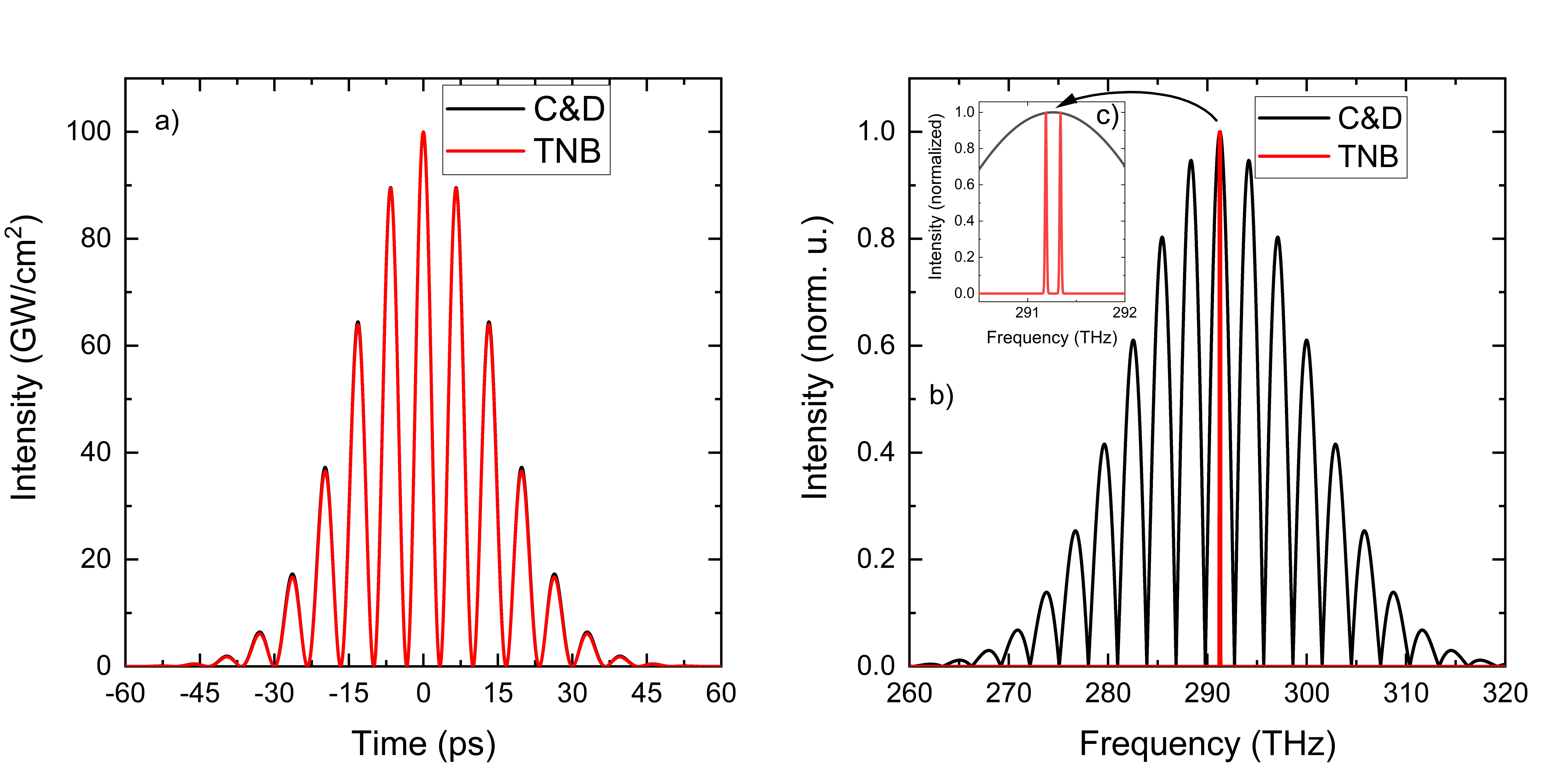}
\caption{Comparison of the C\&D and the TNB techniques: (a) the intensities as a function of time, (b) the spectral intensities.}
\end{figure}

The properties of the THz sources when pumped by the above discussed pump pulses were investigated using the numerical model described in Ref. \cite{Ravi2016}, with the addition of taking into account the Fresnel losses during the exit of the THz pulse from the nonlinear crystal. The periodic structure of the wafer stacks was considered by changing the sign of the nonlinear coefficient at each half-period. The model however does not take into account the transmission losses between the individual wafers.

\section{Results and discussion}

\subsection{Wafer stack structure pumped by a FL or chirped Gaussian pulse}

Terahertz generation by a FL Gaussian-shaped pump pulse (Fig. 1a) with 100 GW/cm$^2$ peak intensity was investigated in a five-period WS structure. The period of the WS structure was chosen to be $\Lambda =0.74$ mm, which (according to Eq. 1) corresponds to a $\nu_0 = 0.15$ THz frequency. The whole length of the five-period structure is 3.7 mm.

Fig. 3a shows the THz-generation efficiency along the pump pulse propagation in the WS structure for a few different pump pulse durations. Fig. 3b shows how the efficiency at the exit from the structure depends on the pump pulse duration. THz pulse shape and its spectra are also shown as insets in Fig. 3b for the cases of three different pump pulse durations: the (near to) maximum efficiency case at 1.5 ps, as well as shorter (0.5 ps) and longer (2.5 ps) pulses. As it can be seen, both THz generation efficiency and the shape of the generated THz pulses significantly depend on the pump pulse duration. The efficiency has a maximum at a specific $\tau_{\mathit{opt}}$ pump pulse duration, which we will call as “optimal pump pulse duration”. This is approximately 1.6 ps for the selected $\nu_0 = 0.15$ THz frequency. 

\begin{figure}[H]
\centering\includegraphics[width=13cm]{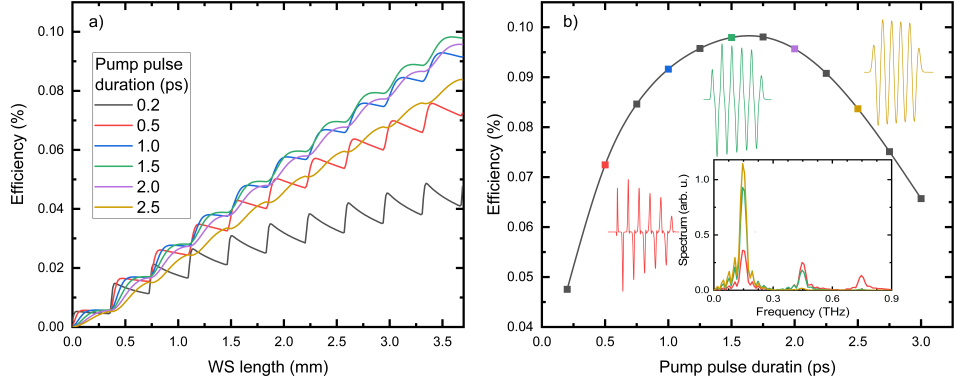}
\caption{Conversion efficiencies in WS for different pump pulse durations. (a) efficiency vs. the WS length, and (b) efficiency at the end of the WS vs. the pump pulse duration. The THz pulse shape and the spectrum (in the inset) are given for three pulse durations identified by colors.}
\end{figure}

It can be clearly seen that spectral peaks appear not only at the desired central frequency, but its odd multiples too, leading to distortions in the waveform. To understand the appearance of these odd harmonics, let’s consider the following approximation for the THz spectral energy density generated by optical rectification:

\begin{equation}
W(\Omega, L) \propto
 \Omega^2 L^2 \tau^2 e^{-\frac{\tau^2 \Omega^2}{8\ln 2}} \mathrm{sinc^2} \left( \frac{\Delta k L}{2}\right),
\end{equation}
which is obtained by neglecting all effects on the pump pulse \cite{Toth2023, Vodopyanov2006}. In this equation, $\Omega$ is the THz angular frequency, $L$ is the thickness of one wafer, and

\begin{equation}
\Delta kL =
 \frac{\Omega}{c}\left( n_{\mathit{THz}}-n_p \right) L
\end{equation}
is the phase mismatch. As long as the $\Delta k = 0$ phase matching condition is fulfilled (in which case the value of the sinc function is 1), the spectral bandwidth is determined by the length of the pump pulse through the $\Omega^2 \exp \left( \frac{-\tau^2 \Omega^2}{8\ln 2}\right)$ term, which peaks at $\nu_0=0.375/\tau$ \cite{Toth2023}. In our case however $\Delta k>0$, so expanding the sinc function, then substituting $\Delta k$, we get

\begin{equation}
W(\Omega, L) \propto
 \frac{\Omega^2 \tau^2}{\Delta k^2} e^{-\frac{\tau^2 \Omega^2}{8 \ln 2}} \sin^2 \left( \frac{\Delta k L}{2}\right) \propto \tau^2 e^{-\frac{\tau^2 \Omega^2}{8\ln 2}} \sin^2 \left( \frac{\Omega L}{2c}(n_{\mathit{THz}}-n_g)\right).
\end{equation}
That is, the maxima are at approximately the maxima of the $\sin^2 (x)$ function, situated at

\begin{equation}
\Omega_m =
\frac{(1+2m)\pi c}{(n_{\mathit{THz}}-n_g)L}
\end{equation}
angular frequencies. Since the period of the WS is $\Lambda=2L,$ from Eq. 14 the THz spectrum has maxima at the

\begin{equation}
\nu_m =
\frac{c}{\Lambda (n_{\mathit{THz}}-n_g)}(1+2m)
\end{equation}
frequencies, where the $m=0$ case gives back Eq. 1, while the $m>0$ explains the appearance of the odd harmonics.

The inset of Fig. 3b also shows that for shorter pump pulse lengths, the odd harmonics are much more significant. For the 500 fs long pump (red line), even a fourth spectral peak is significant, while for the 2.5 ps one (yellow line) practically only the main spectral component is visible. The faster decrease of the spectral peaks with frequency for longer pulse duration is explained by the exponential term in Eq. 13. This is expected, since longer pump pulses don’t have a wide enough spectrum for higher frequency peaks.

The THz generation efficiency was investigated at the approximately optimal (1.5 ps) pulse duration for different peak intensities. Fig. 4a shows the THz-generation efficiency along the pump pulse propagation in the WS structure at different intensities. Fig. 4b shows how the efficiency achieved with a five-period WS structure depends on the pump intensity. For comparison a linear fitting curve (red line) is also displayed. The results of the numerical calculation (black squares) are sitting perfectly on the red line, demonstrating a linear dependence of the efficiency on the pump intensity. The efficiency also linearly depends on the crystal length (number of wafer pairs) for all pump intensities (see Fig. 4a). This simple dependence is expected supposing that the self-phase modulation (SPM) and the cascade effects (CE) \cite{Ravi2014} do not significantly influence the THz-generation process. Indeed, we did not observe significant change in the efficiency when either SPM and/or CE were switched on and off, even though the spectrum of the pump pulse significantly changed during the propagation inside the WS.

THz-generation efficiency for WS structures with up to 20 periods at various frequencies was also studied, the results of which are given in Fig. S1 of the Supplementary Material. Based on these results, it can be stated that the lower the THz frequency, the longer will the linearity between the efficiency and the number of wafer periods hold. At higher frequencies, the deviation from this linear behavior is caused by the significant THz absorption.

\begin{figure}[H]
\centering\includegraphics[width=13cm]{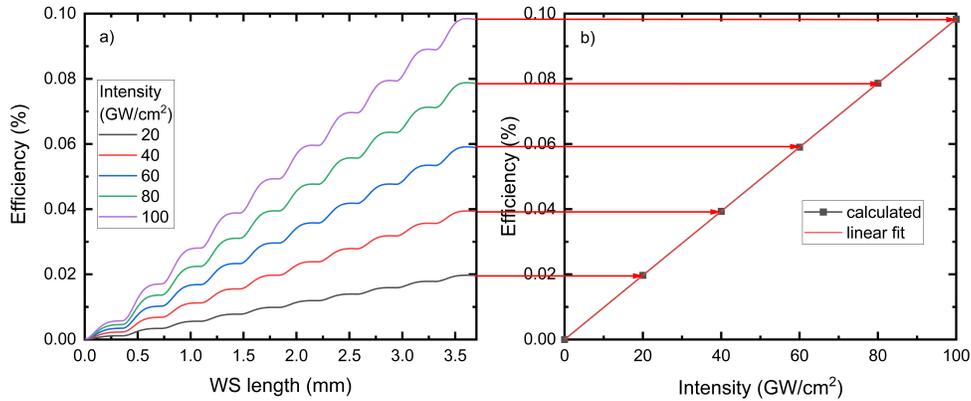}
\caption{Conversion efficiencies for different pump intensities in a 5 period WS. (a) Efficiency vs. the propagation length along the WS structure, (b) efficiency at the end of the WS vs. the pump intensity.}
\end{figure}

Assuming a five-period WS structures with a few different period lengths corresponding to different desired THz frequencies on the 0.1-2.0 THz range, we calculated (similarly as illustrated in Fig. 3b for 0.15 THz) the $\tau_{\mathit{opt}}$ optimal pump pulse duration. The obtained $\tau_{\mathit{opt}}$ values are indicated by blue dots in Fig. 5. As can be seen, these dots fit very well to a 

\begin{equation}
\tau_{\mathit{opt}} =
\frac{0.245}{\nu_0}=0.245\cdot T_0
\end{equation}
curve, where $T_0$ is the temporal period of the generated THz radiation. This relationship can be used for maximizing the THz generation efficiency in the case of LN WS THz sources working in the 0.1-2.0 THz spectral range. It is worth to note that this inverse proportional relation between the optimal pump pulse duration and the THz frequency was noticed earlier \cite{Ravi2016}. Furthermore, in an early paper \cite{Lee2002} on PPLN THz sources it was shown, based on a simple numerical model, that the maximum THz amplitude can be achieved when 
\begin{equation}
\frac{d}{d_w} =
2.2,
\end{equation}
where $d=\Lambda /2$ is the domain width, in our case the wafer thickness, and 
\begin{equation}
d_w =
\frac{c\tau_m}{n_{\mathit{THz}}^{ph}-n_{p}^{gr}}
\end{equation}
is the walk-off length for the $\tau_m$ pump pulse duration.

Expressing $\tau_m$ from Eq. 18 and applying Eq. 1 and Eq. 17, we can arrive to
\begin{equation}
\tau_m =
\frac{1}{4.4\nu_0}=0.227\cdot T_0.
\end{equation}
The green line in Fig. 5 shows this $\tau_m (\nu_0)$ dependence. The blue dots corresponding to the optimal pump pulse duration calculated by our numerical model are situated above this green line.

\begin{figure}[H]
\centering\includegraphics[width=13cm]{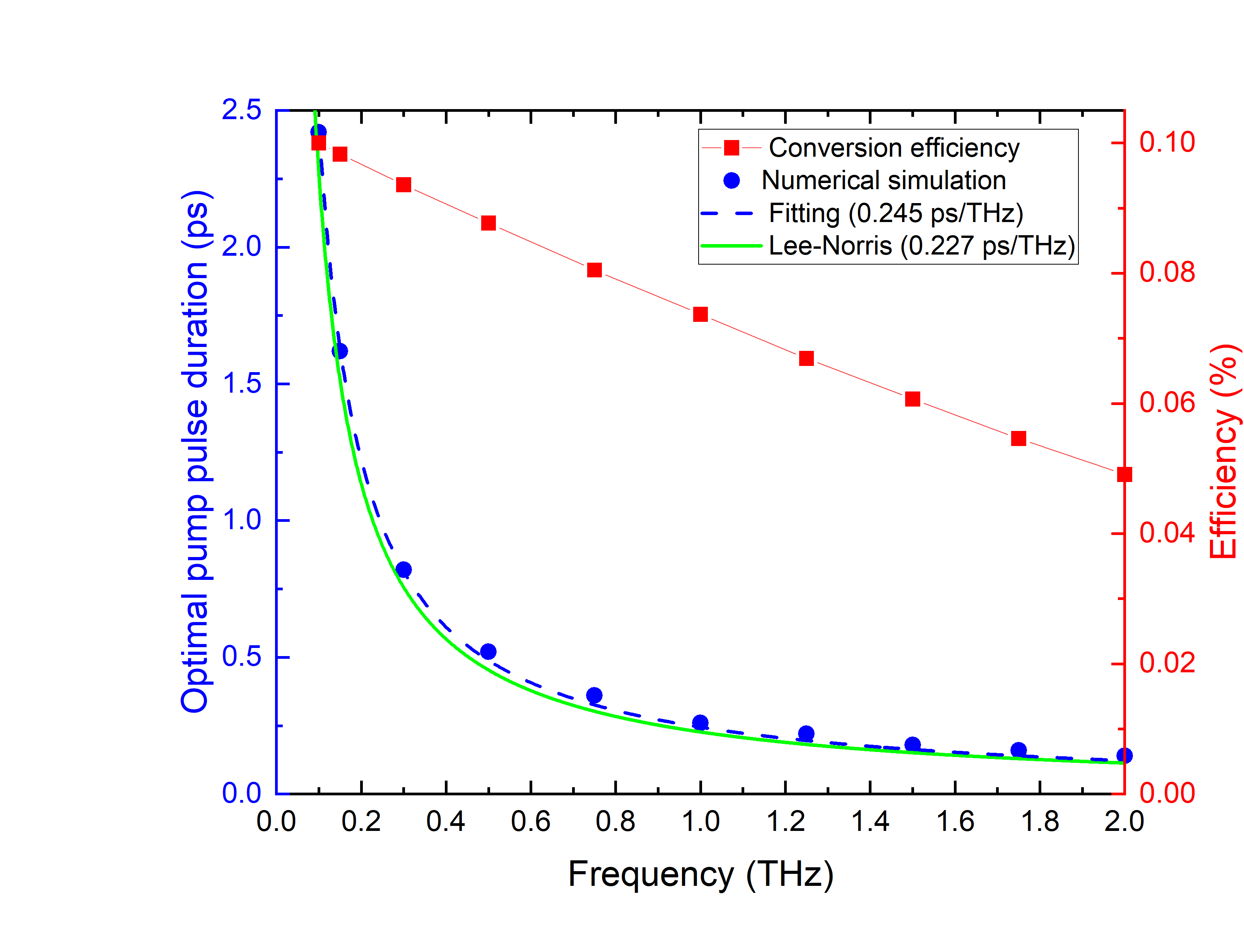}
\caption{Frequency dependance of the conversion efficiency and optimal pump pulse duration. The red connected squares plot the maximum efficiency achieved for the given frequencies. The pump pulse duration for which this maximum is achieved is given in blue dots. The dashed blue line shows an inverse proportional function fit to these dots, the formula of which is $\tau_{\mathit{opt}}=0.245/\nu$. The green line also shows an inverse proportional function, where the 0.227 coefficient has been taken from \cite{Lee2002}.}
\end{figure}

Fig. 5 also shows the THz-generation efficiency as the function of the THz frequency. About a factor of two decrease of the efficiency can be observed when the frequency is increased from 0.1 THz to 2.0 THz. To explain this, it is important to underline that during the calculations the number of periods and not the total length of the crystal was kept constant; for example, for a 1.5 THz central frequency the length of the 5-period crystal is 0.37 mm, while for a 0.15 THz one it is 3.7 mm. That is, if there was no absorption, changing the THz frequency and inverse proportionally the pump pulse length would not lead to significant changes in the main efficiency, since it is proportional to the product of the frequency and the crystal length (see Eq. 11, for example). Even in the presence of absorption, we could expect an almost frequency independent efficiency if the absorption coefficient would be proportional to the frequency, since the effect of absorption depends on the product of the absorption coefficient and the crystal length. In reality, however, the THz absorption strongly increases with the frequency; on the investigated range, it is proportional to the square of the frequency. As a result, at 2.0 THz the transmission of the five period WS is only 0.34 times the transmission at 0.1 THz, and the transmission from the middle of the WS to its end at 2.0 THz is only 0.58 times the transmission at 0.1 THz. These numbers explain the about two times decrease of the efficiency when the frequency is increased from 0.1 THz to 2.0 THz, as shown in Fig. 5.

Calculations have also been performed where pulses having broad spectral band supporting 0.2 ps FL pulse duration were chirped (the case of Fig. 1d) to the same length (0.5 - 2.5 ps) and intensity as for the FL ones previously shown. The results showed no significant difference between the two cases (see Fig. S2, S3 in the Supplementary Material). This means that broadband lasers provide a possibility to maximizethe THz generation efficiency by adjusting their pulse duration.

\subsection{Wafer stack structure pumped by a pulse sequence}
To achieve higher conversion efficiency in WS, the use of pulse sequence pumping (see Fig. S4 in the Supplementary Material) (Fig. 1b or Fig. 1e) or modulated intensity pulse (Fig. 1c or Fig. 1f) instead of a single pulse is required. At this point, we note that the results obtained for sequences of FL (Fig. 1b) and of chirped pulses (Fig. 1e) were, to a great extent, the same.

By pumping a WS consisting of $N_{WS}$ periods with a pulse sequence consisting of $N_p$ pulses, the cycle number of the resulting THz radiation is determined by the following simple formula:
\begin{equation}
N_{\mathit{cycl}} =
N_{WS}+N_p-1.
\end{equation}
At first, the case of a given WS pumped by a sequence of an increasing number of pulses was investigated. The number of THz cycles increased according to Eq. 20. Then, the case of a given number of desired THz cycles was investigated, where this number was achieved by varying both the number of wafer pairs and pump pulses according to Eq. 20.

Fig. 6.a shows the dependence of the THz generation efficiency on the number of pump pulses, $N_p$, in WS having 5 and 7 periods. For a small number of pump pulses, the efficiency increases rapidly, while for larger number of pump pulses, it tends to a maximum. This behavior can be explained with the following simple model: all individual pump pulses of the pulse sequence generate a THz pulse having $N_{WS}$ periods with $E_0$ amplitude. However, these THz pulses are shifted relative to each other by an integer multiple of the time period. Thus, the amplitude in the middle of the resulting THz pulse will be $N_p\cdot E_0$ (until $N_p \leq N_{WS}$). However, moving towards the beginning or the end of the resulted pulse the amplitude will be $(N_p-1)\cdot E_0,~(N_p-2)\cdot E_0,$ etc. Considering that the energy of the resulted THz pulse is proportional to the sum of the squares of the amplitudes belonging to each period, and the energy of the pumping pulse sequence is $N_p$ times the energy of one pulse, we obtain the following expression for the efficiency:
\begin{equation}
\eta =\frac{W_{\mathit{THz}}}{W_{PS}} \approx \begin{cases}
\dfrac{3N_{WS}N_p-N_p^2+1}{3}~\cdot \eta_0, &\text{if $N_p \leq N_{WS}$},\\
\dfrac{N_{WS}(3N_{WS}N_p-N_{WS}^2+1)}{3N_p}~\cdot \eta_0, &\text{if $N_p \geq N_{WS}$}.
\end{cases}
\end{equation}
where $W_{PS}$ and $W_{\mathit{THz}}$ are the energy of the pulse sequence and the THz pulse generated by the WS, and $\eta_0$ is the generation efficiency of one wafer pair pumped by one single pulse. According to Eq. 21, for $N_p \gg N_{WS}$, the efficiency approaches $N_{WS}^2\cdot \eta_0$, that is $N_{WS}$ times larger than the efficiency of the WS in the case of using single pulse pumping. It should be noted that even though this is a simplistic model that does not take into account e.g. THz absorption and dispersion, the results obtained with it (lines in Fig. 6a) fit the simulated results (squares) quite well. 

\begin{figure}[H]
\centering\includegraphics[width=13cm]{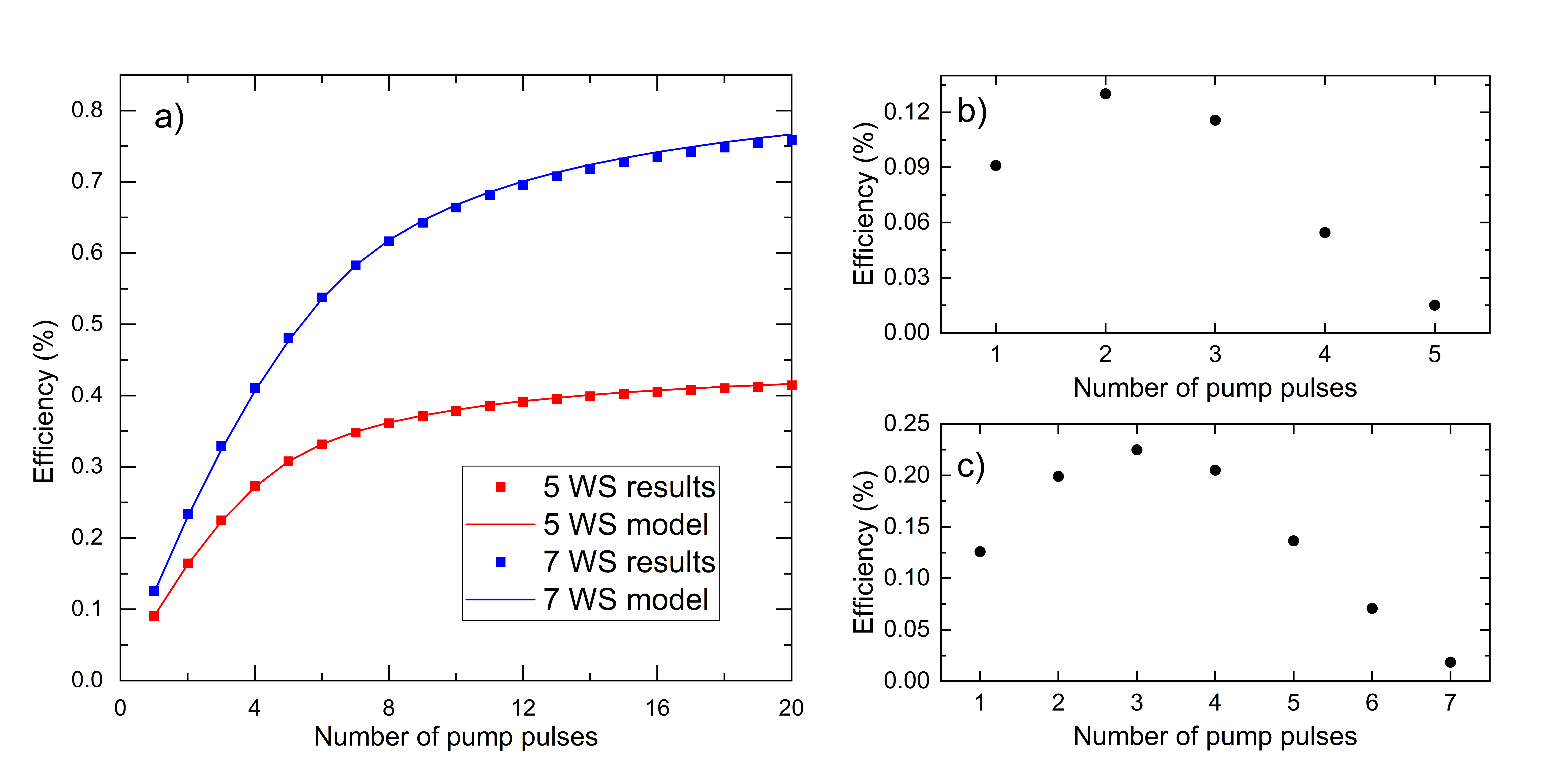}
\caption{Conversion efficiency dependence on the number of pump pulses. a)  The number of WS periods is fixed to 5 and 7. Apart from simulation results (depicted by squares) the results of the simple model described are given (depicted by lines). b, c) the number of WS periods and pump pulses is varied so that always 5 or 7 THz cycles are generated.}
\end{figure}

Fig. 6b and 6c shows the efficiency dependence on the pump pulse number when the goal is to generate 5-cycle and 7-cycle THz pulses, respectively. According to our calculations, the highest conversion efficiency can be achieved with a configuration such that the number of pulses is equal to the number of stack periods.  In cases of odd number of cycles, $N_{WS} > N_p$ should be used.

At this point, we would like to mention the Gires-Tournois etalon as a very practical device to generate pulse sequences (albeit with varying intensity) from a single pulse \cite{Yeh2008, Nasi2022}. In this case, the FL Gaussian pulse goes through the inner structure of a plane-parallel substrate having a partial reflecting layer on its front surface, and a layer having 100\% reflection on its back surface.  Reflecting a pulse on such an etalon, a pulse sequence is generated. Choosing 38\% reflection for the front layer, the intensity of the first two pulses will be equal \cite{Nasi2022, GT}.

\subsection{Wafer stack structure pumped by intensity-modulated pulse}

 In practice it is more common to generate an intensity modulated pulse (Fig. 1c, or Fig. 1f) than pulse sequences. It is important to note that there are two crucial differences between intensity-modulated pulses and pulse sequences. First, pulse sequences (which are considered in this article) contain identical pulses with the same peak intensity, in contrast to the intensity modulated pulse, which has varying peak intensity. Second, and more important, in the case of pulse sequences the pulse widths of the individual pulses are independent from the tracking distance of the pulses. In contrary, in the case of intensity-modulated pulses, up to a good approximation, the width of the individual pulses is half the tracking distance. 

To compare the results of THz generation using pulse sequence and intensity-modulated pulse for pumping, four different pump pulses were assumed as presented in Fig. 7a. The two intensity-modulated pulses (created by the TNB and C\&D techniques) were chosen so that the modulation frequency be 0.15 THz and the intensity FWHM of the whole pulse be 33.3 ps. The pulse duration of the individual pulses for one sequence was 3 ps (about the same as the duration of one peak of the intensity-modulated pulses) and 1.5 ps, the (close to) optimal pump pulse duration for the other. Fig. 7b shows the conversion efficiencies during propagation in a 10 period WS for all four pump cases. The conversion efficiency in the case of both types of intensity-modulated pump pulses and the pulse sequence with 3 ps pulse duration increases at the same rate in the WS. Intensity-modulated pump pulses created either by the TNB or the C\&D techniques result in not only the same efficiency, but the same THz pulse shapes too (see Fig. S5, S6 in the Supplementary Material). However, significantly higher efficiency can be achieved with a pulse sequence consisting of 1.5 ps duration pulses. Thus, in terms of efficiency, using a pulse sequence with the optimal pump pulse duration can be more advantageous than using intensity-modulated pulses. 

\begin{figure}[H]
\centering\includegraphics[width=13cm]{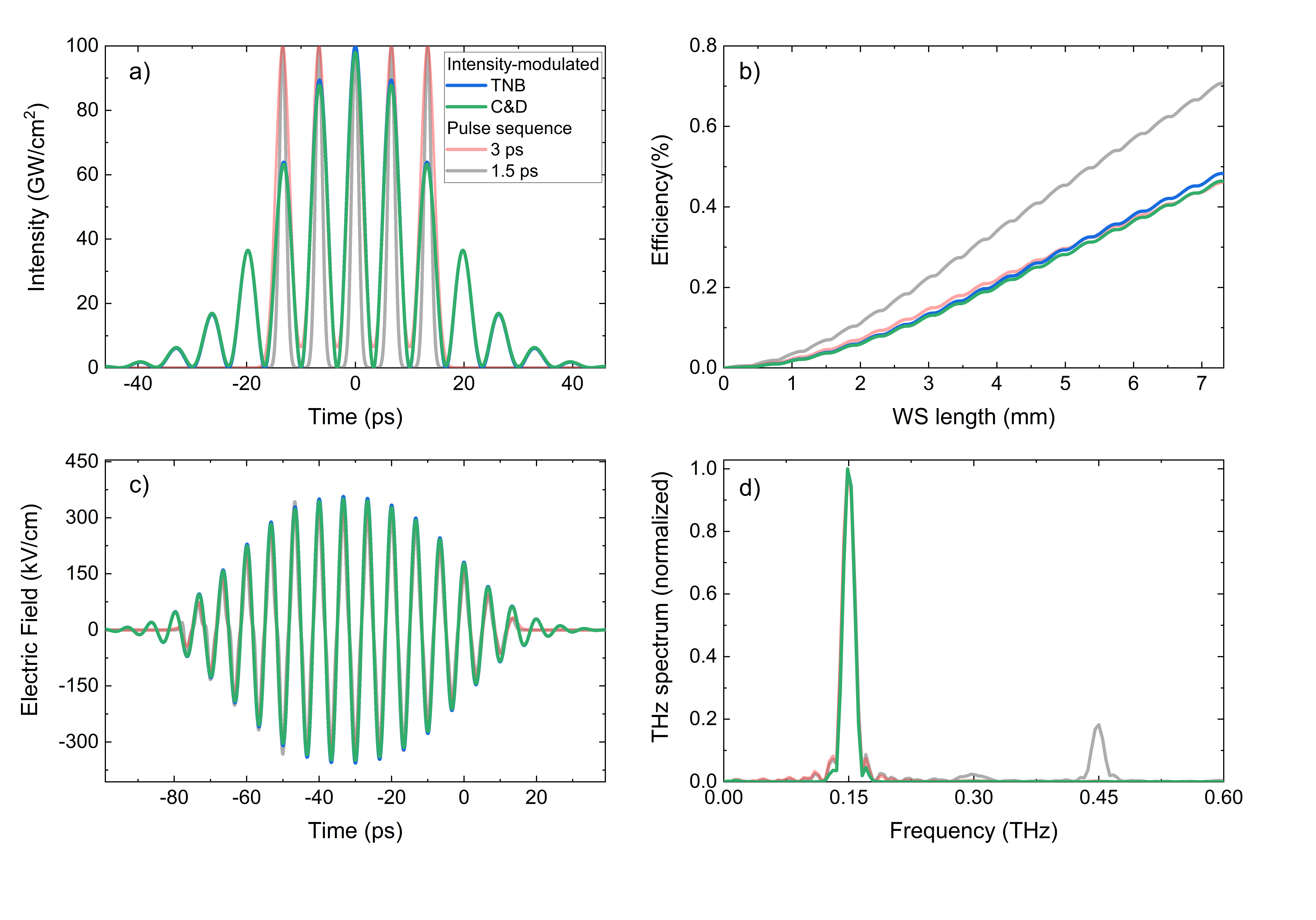}
\caption{(a) Intensity shapes of the pump pulses: intensity modulated TNB (blue) and C\&D (green) pulses, as well as pulse sequences with 1.5 ps (grey) and 3 ps (red) individual pulse durations with 6.67 ps tracking distance between them; (b) THz conversion efficiencies, (c) generated THz waveforms and (d) their spectra for all investigated cases.}
\end{figure}

Figures 7c and 7d show the temporal shapes and the spectra of the THz fields generated by the ten period WS pumped by the mentioned four different pump pulses. As can be seen, the resulting THz field shapes are almost exactly the same. The main peaks of the four spectra are also very similar to each other.  The spectrum corresponding to the case of pumping with the optimal (1.5 ps) pump pulse duration contains however a significant third harmonic component. This component is missing in the other three cases, where the duration of the pulses is two times longer (3.0 ps). The reason for this behavior is the same as was discussed in connection to the inset of Fig. 3b.

\subsection{Multicycle THz pulse generation by tilted pulse front}

Multicycle THz pulse generation is also possible with TPFP setups if pulse sequences or intensity-modulated pump pulses are used. In this chapter, such a possibility is investigated for pulses created using the C\&D and TNB techniques. The intensities of the pump pulses having 0.15 THz modulation frequency are shown in Fig. 8a. Similar to Fig. 7a, the curves corresponding to the C\&D and TNB techniques are perfectly overlapping irrespective of the (short enough) FL pulse duration in the C\&D technique. 

\begin{figure}[H]
\centering\includegraphics[width=13cm]{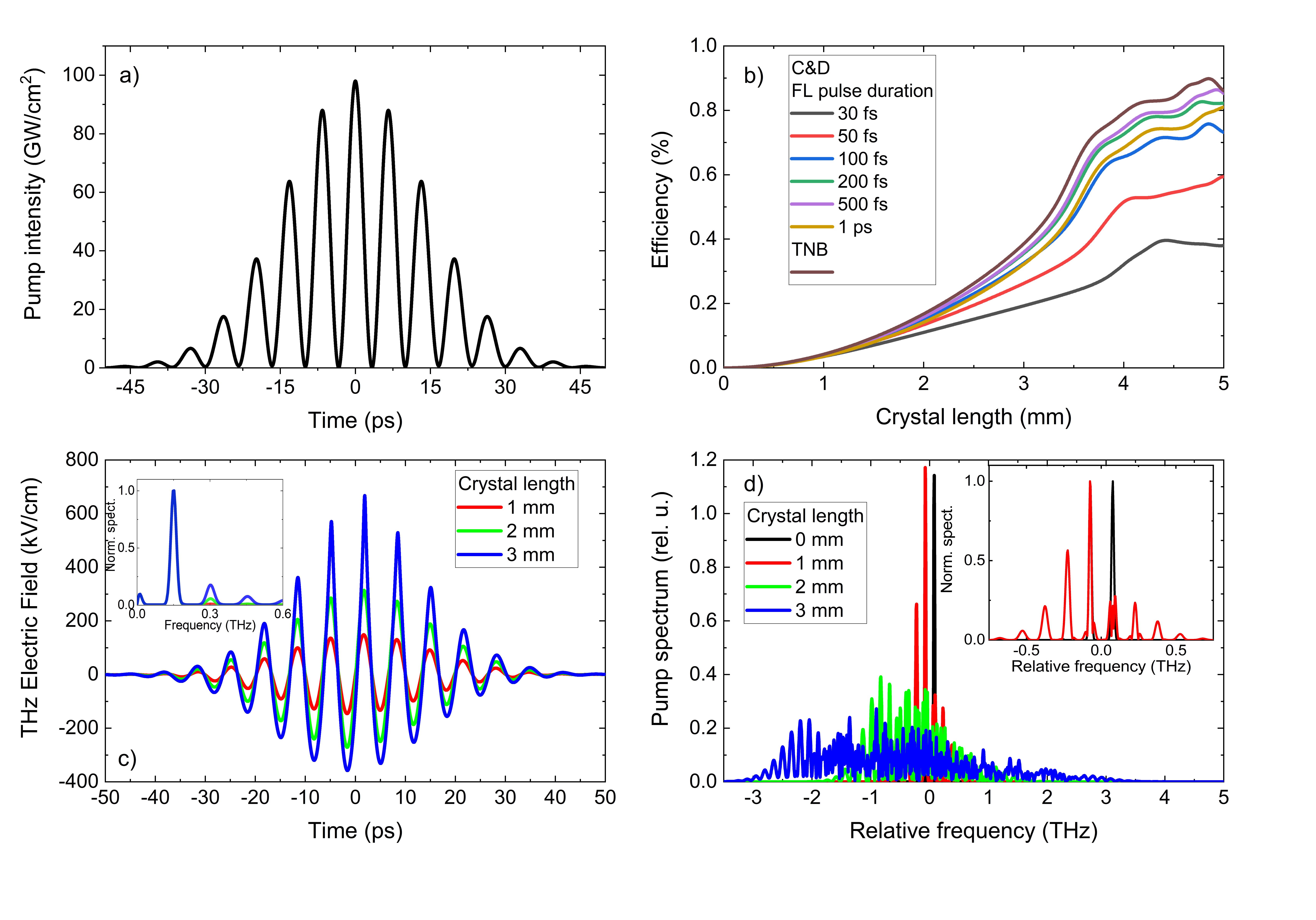}
\caption{Results for multicycle THz pulse generation in a TPFP setup using intensity-modulated pump pulses. (a) Temporal intensity profiles of the pump pulses, (b) Conversion efficiency vs. the propagation distance inside the crystal for different FL durations in the C\&D technique as well as in the TNB case, (c) THz electric fields and amplitude spectra (inset) at different propagation distances inside the crystal for the TNB case, (d) Spectra of the TNB pump pulse at different propagation distances in the LN crystal.}
\end{figure}

In Fig. 8b the THz conversion efficiency is given versus the propagation length inside the lithium niobate crystal for the C\&D (with different FL pulse duration) and TNB cases, assuming 100 GW/cm$^2$ pump intensity. As can be seen, the TNB technique results in higher efficiency than the C\&D for most of the FL pulse lengths. Furthermore, in the case of C\&D the efficiency strongly depends on the FL laser pulse duration, despite the same temporal shape (see Fig. 8a) of the modulated pulse.  The highest efficiency can be reached for an about 500 fs FL pump pulse duration. In this case the efficiency approaches the one for TNB pump pulses. For shorter FL pulse duration, the efficiency decreases strongly. It is important to note that this phenomenon only exists in the TPFP, however not in the WS setup. In the TPFP case the reason of the decreasing efficiency with decreasing FL pulse duration (compared to the optimum 500 fs) is the large GDD caused by the angular dispersion introduced via the pump pulse front tilt \cite{Hebling1996, Fueloep2010}. For pump pulses longer than the optimum, the efficiency also drops. The reason behind this is the decreasing amplitudes of the high frequency components in the pump.

The THz waveform in the cases of 1-, 2- and 3 mm crystal lengths for the TNB technique are shown in Fig. 8c. While for the 1- and 2 mm lengths the THz waveform is near sinusoidal, for 3 mm length the electric field peaks are clearly distorted, and the spectrum contains significant peaks at higher harmonics (see inset). For crystal lengths above 3 mm, the efficiency does not increase monotonically anymore, and the THz waveform becomes drastically distorted (see Fig. S7, S8 in the Supplementary Material).

Fig. 8d shows the pump spectrum (shifted by the central frequency of the pump) in the case of pumping by TNB pulse at different crystal lengths. While at the entrance only two components with equal amplitudes are present with a frequency difference of 0.15 THz (see the inset), new peaks with significant amplitudes appear already at the 1 mm crystal length, (much earlier than the distortion of the THz waveform become visible, see Fig. 8c), again at 0.15 THz from each other, due to the cascade effect \cite{Ravi2016}. The new peaks are stronger at the low frequency side of the original peaks than at the high frequency side, resulting a downshift of the average spectrum. This continues for longer crystal lengths and components with smaller frequency differences appears, resulting a chaotic, quasi-continuous spectrum (see the curves for 2- and 3 mm in the main Fig. 8d).

Using pump pulses created by C\&D technique with optimal FL pulse duration (0.5 ps for 0.15 THz generation, see Fig. 8b), THz pulses with the same temporal shape and spectrum are generated than using TNB pumping. For crystal length – pump intensity combinations resulting high efficiency, the generated THz pulses are single-signed like (not symmetrical in the electric field axis), and the spectrum contains a very large number of harmonics (see Supplementary Material Fig. S8). Such close to sine-like THz pulses as are shown in the inset of Fig. 3b, and in Fig. 8c can be generated for crystal length – pump intensity combination resulting less than a 0.2\% efficiency.   

Calculations have also been performed for TPFP pumped by a pulse sequence. The tracking distance was 6.67 ps, corresponding to the 0.15 THz frequency requirement. The pulses had a temporal lengths of 1 ps, 2 ps and 3 ps, as shown in Fig. 9b. The THz generation efficiency is plotted in Fig. 9a, which shows that generation efficiencies exceeding 1\% are achievable. However, similarly to the C\&D and the TNB techniques, with a pulse sequence the sinusoidal character of the generated THz pulse deteriorates once the generation efficiency rises above 0.2\%.

\begin{figure}[H]
\centering\includegraphics[width=13cm]{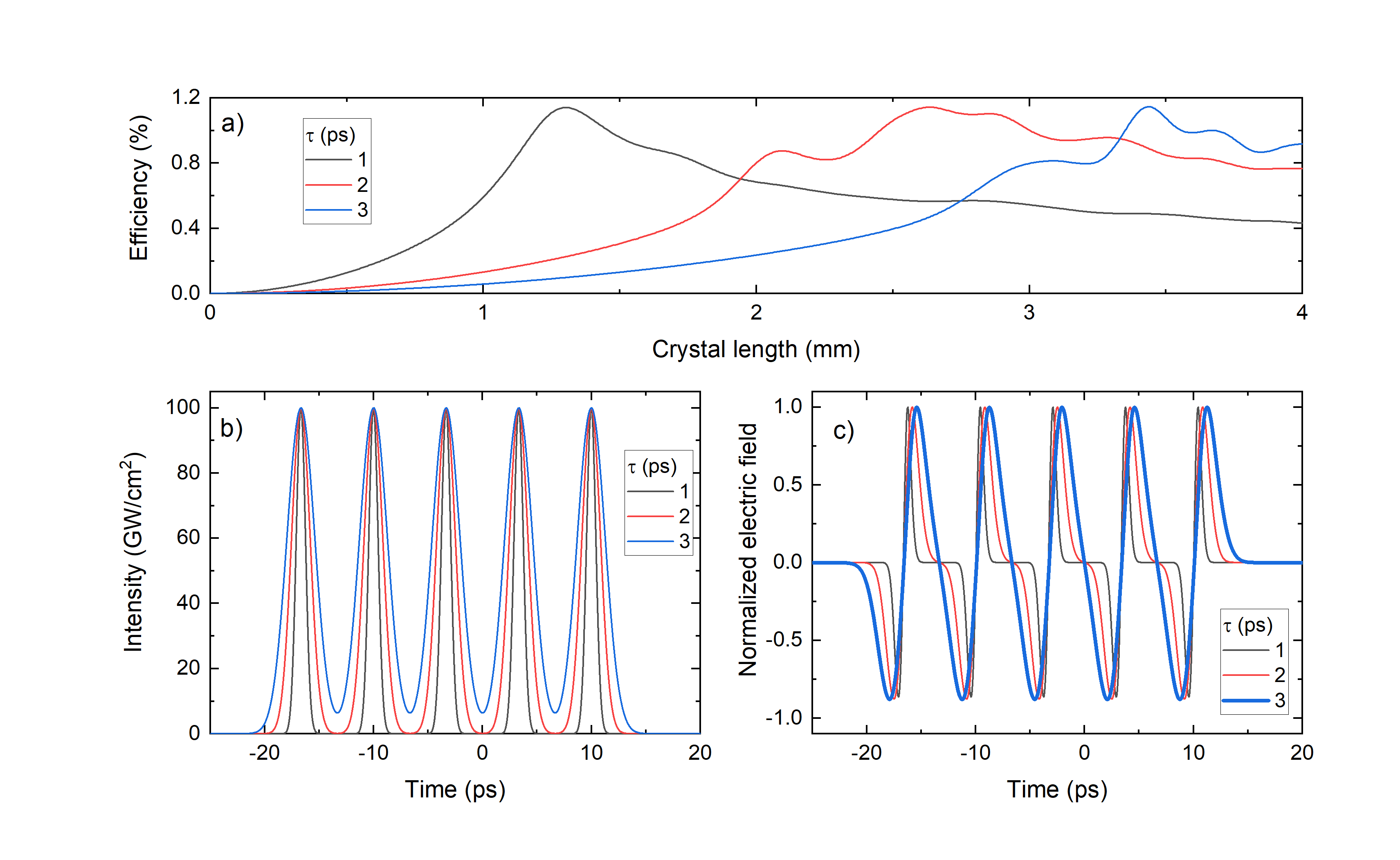}
\caption{Results of the simulation considering TPFP pumped by a pulse sequence containing 5 pulses with lengths of 1 ps (black), 2 ps (red) and 3 ps (blue). (a) Efficiency vs. the crystal length, (b) the temporal profile of the pump pulses, (c) the generated THz fields for conditions resulting efficiencies below 0.2\%.}
\end{figure}

In Fig. 9c, we show the shapes of the generated THz pulses for different pulse durations of the pulse sequence in cases where the generation efficiency is still below 0.2\%. It can be seen that with 1 ps pumping, a THz pulse sequence instead of a continuous multicycle THz waveform is obtained. Even with 2 ps pumping, the regime where the THz profile is sinusoidal is still not achieved. For a sinusoidal THz field in the case of a pulse sequence, the temporal length of each pulse should be approximately half of the tracking distance.

In Fig. 10 the THz spectra obtained by all the methods (except the pulse sequence pumped TPFP) considered are given. It can be seen that for TPFP and pulse sequence pumped WS, the main spectral component is narrower and its wings are much smaller than for the other cases. Both for one-pulse and pulse sequence pumping of WS the third harmonic appears with a similarly large spectral intensity. Pumping the WS by an intensity-modulated pulse yields a main component of slightly larger FWHM, while the wings and third harmonics will be smaller.

It is important to emphasize that for intensity-modulated pumping of the TPFP setup, contrary to WS, the bandwidth of the pump pulses does indeed affect the THz generation. The reason behind this is the large GDD caused by the tilted pulse front. As a result, in the C\&D technique, THz generation strongly depends on the FL pulse duration. 

\begin{figure}[H]
\centering\includegraphics[width=8cm]{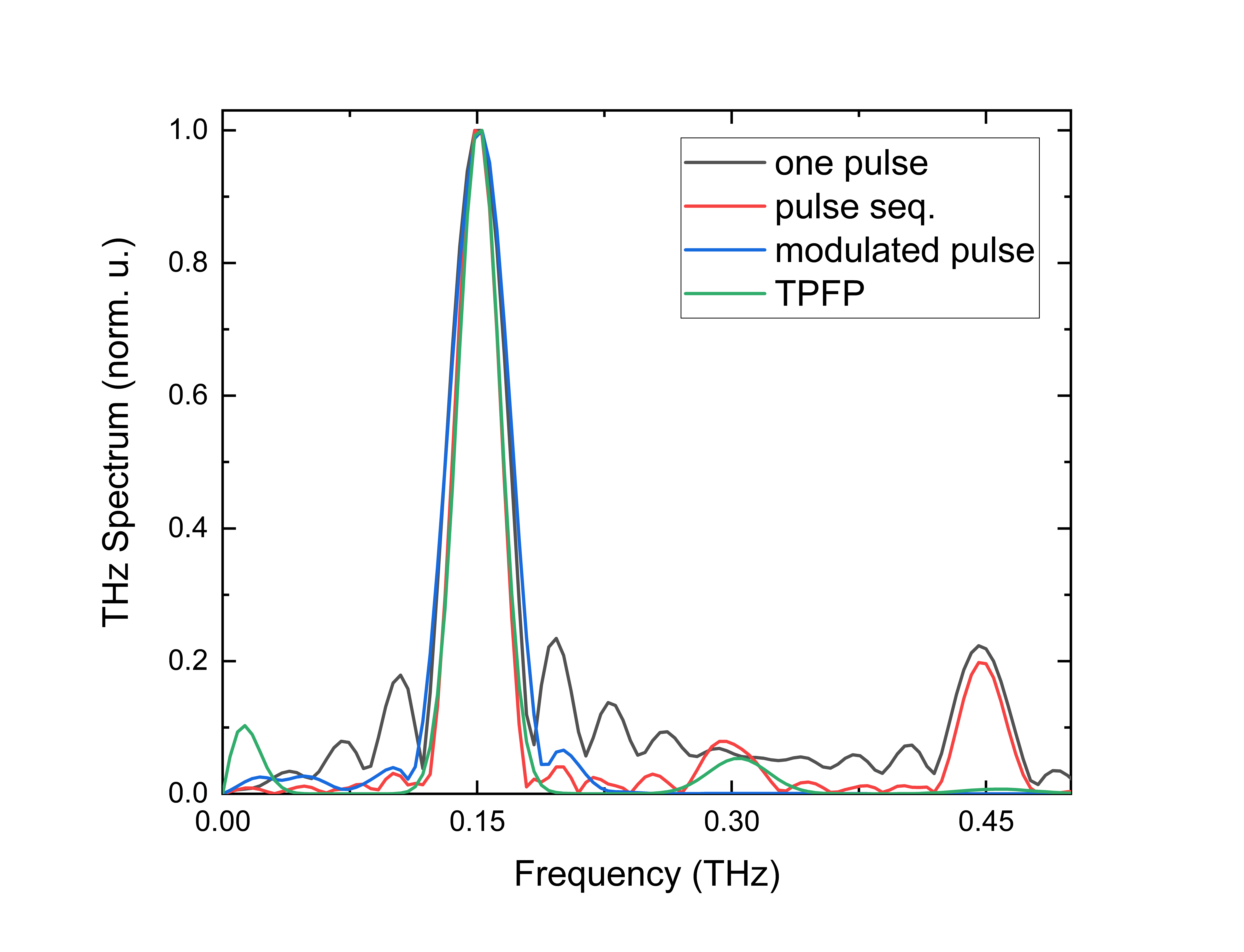}
\caption{THz spectra achieved with different techniques: WS pumped by single pulse (black), by pulse sequence (red), and intensity-modulated pump (blue), as well as TPFP setup pumped by intensity-modulated pulse (green).}
\end{figure}

Similar to the case of WS, THz pulse generation at higher frequencies is also possible using TPFP setup. Furthermore, contrary to the case of WS, using TPFP with intensity-modulated pulses generated by the C\&D technique allows simple adjustments of the THz frequency by changing the pump pulses (changing the delay between the two chirped pulses). On the 1.0 – 2.0 THz range a small adjustment of the setup is also needed, while on the < 1 THz range the frequency tuning is possible without any adjustment of the TPFP setup.

\subsection{The effect of the pumping beam profile on the efficiency}

In the case of WS, as well as those TPFP THz sources which have plan-parallel nonlinear optical material slab \cite{Toth2023}, the fact that the efficiency is proportional to the pump intensity in the case of 1D model (see Fig. 4) provides an opportunity to give an estimation to the THz conversion efficiency for pump beams having known transversal intensity profile.

If the pump beam is axially symmetric and the beam size is large enough, thus making it possible to neglect the diffraction inside the crystal, the THz efficiency for the beam can be calculated as:
\begin{equation}
\eta_b =
\frac{W_{\mathit{THz}}}{W_p}=\frac{\int\limits_0^{\infty}\eta (I(r))I(r)r\dd r}{\int\limits_0^{\infty}I(r)r\dd r},
\end{equation}
where $r$ is the radial coordinate, $I(r)$ is the $r$ dependent intensity (the transversal beam profile),  and $\eta(I(r))$ is the intensity-dependent THz conversion efficiency based on the 1D model. Because the $\eta(I(r))$ efficiency is proportional to the intensity, it can be written as

\begin{equation}
\eta (I(r)) =
\alpha_{\eta}I(r),
\end{equation}
and Eq. 22 can be reduced to

\begin{equation}
\eta_b =
\alpha_{\eta}~\frac{\int\limits_0^{\infty} I^2(r)r\dd r}{\int\limits_0^{\infty}I(r)r\dd r}.
\end{equation}
While the beam intensity profile of laser oscillators or small energy laser amplifiers can be well described by a Gaussian distribution, the intensity beam profile of high energy amplifiers usually approaches a super-Gaussian distribution according to

\begin{equation}
I(r) =
I_0e^{-2\frac{r^{2p}}{w^{2p}}},
\end{equation}
where $w$ is the beam radius at $1/e^2$ intensity of the maximum. For $p = 1$ Eq. 25. describes the Gaussian distribution, while for $p \to \infty$ it describes a flat-top beam. Fig. 11a shows the distributions for a few different $p$ cases. 

To obtain the efficiency of the pump beam for different $p-$values, Eq. 25 can be substituted into Eq. 24, which gives

\begin{equation}
\eta_b =
\alpha_{\eta}~\frac{I_0^2\int\limits_0^{\infty} re^{-4\frac{r^{2p}}{w^{2p}}}\dd r}{I_0\int\limits_0^{\infty}re^{-2\frac{r^{2p}}{w^{2p}}}\dd r}=\alpha_{\eta} I_0\frac{\frac{\Gamma \left( 1/p \right)}{2p2^{\frac{2}{p}}}w^2}{\frac{\Gamma \left( 1/p \right)}{2p2^{\frac{1}{p}}}w^2}=\alpha_{\eta}I_02^{-\frac{1}{p}},
\end{equation}
where the well-known $\Gamma (x)$ function \cite{bronshtein2015handbook} appears, but then drops out. Applying Eq. 23, the beam efficiency in the case of different $p-$ values is

\begin{equation}
\eta_b =
\eta \cdot 2^{-\frac{1}{p}}.
\end{equation}
Figure 11b shows the 2$^{-1/p}$ function. Our results mean that using a pump beam with a Gaussian intensity profile, the efficiency of the multicycle THz-generation in a WS structure is half of what the 1D model predicts. On the other hand, using a flat-top beam the beam efficiency is equal to the efficiency given by the 1D calculation.

\begin{figure}[H]
\centering\includegraphics[width=13cm]{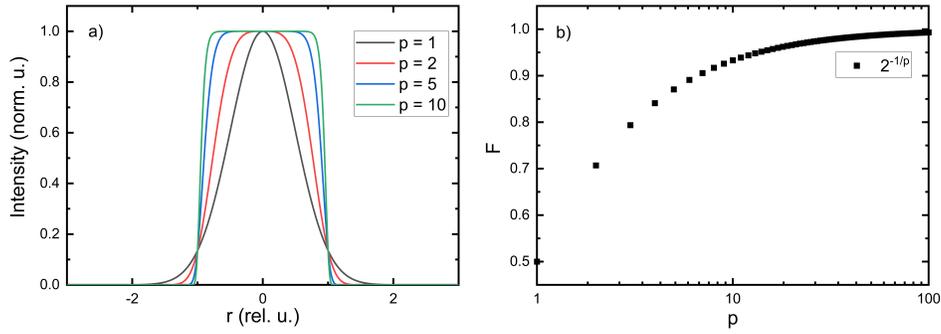}
\caption{(a) Distribution of the pump intensity along the radial direction for different p values. (b) Factor between the THz generation efficiency calculated with the 1D model and the beam efficiency at different $p-$values.}
\end{figure}

Based on Eq. 27, it is possible to determine the dependence of the generated THz pulse energy on the pump energy for a given $w$ beam size and different $p-$values. The pump energy can be calculated as:

\begin{equation}
W_p =
2\pi \cdot I_0\int\limits_{-\infty}^{\infty}\int\limits_0^{\infty} re^{-2\frac{r^{2p}}{w^{2p}}}f(t)\dd r\dd t=2\pi \cdot I_0 \frac{\Gamma \left( 1/p \right)}{2p}2^{-\frac{1}{p}}w^2\int\limits_{-\infty}^{\infty} f(t)\dd t,
\end{equation}
where $f(t)$ is the normalized temporal shape of the intensity; in our case, simply equations 3 to 8 divided by $I_0$. For Gaussian temporal shape, for example,the pump energy will be equal to $W_p=\frac{\pi^{3/2}\Gamma (1/p)}{2\sqrt{2\ln 2}}2^{-\frac{1}{p}}w^2\tau I_0$. According to Eq. 22 the THz energy is given by:

\begin{equation}
W_{\mathit{THz}} =
\eta_b W_p = \alpha_{\eta} I_02^{-\frac{1}{p}}W_p.
\end{equation}
Expressing $I_0$ from Eq. 28 and substituting it into Eq. 29, 

\begin{equation}
W_{\mathit{THz}} =
\frac{\alpha_{\eta}}{\pi w^2}\frac{p}{\Gamma \left( 1/p \right) \int\limits_{-\infty}^{\infty} f(t)\dd t}W_p^2.
\end{equation}
As an example, applying this result to the case of a five-period WS pumped by a single pulse with a 1.5 ps pulse duration from the slope of the intensity-efficiency line in Fig. 4b , $\alpha_{\eta}=\frac{0.098\%}{\mathrm{100~GW/cm^2}}.$ Assuming a $w=$ 5 mm beam waist, Fig. 12 shows the generated THz energy versus the pump energy for different $p-$ values. The points where the peak intensity reaches 100 GW/cm$^2$ for the different $p-$ values has been marked by dots. The curve in the case of Gaussian $(p=1)$ intensity profile overlaps the perfectly flat-top $(p=\infty)$ one, the difference between the two curves being that for all pump energy the peak intensity $(I_0)$ is two times higher in case of Gaussian beam than in case of flat-top profile. The THz energy grows the fastest in the case of a simple super Gaussian $(p=2)$ intensity profile, while the slowest in the case of a Gaussian and flat-top pulse. The rate of growth is determined by the $p-$ dependent part of Eq. 30, which is depicted in the inset of Fig. 12. In the case of a Gaussian beam, the $p-$dependent part is equal to 1, and in the case of $p=2$, it reaches its maximum of 1.13. In the case of a higher $p-$value, the function value tends to 1. 

It is important to note that the presented example was obtained for a single pump pulse in a 5-period WS. The efficiency can be significantly increased using a pump pulse sequence of intensity modulated pump pulses, or/and increasing the number of periods in the WS.

\begin{figure}[H]
\centering\includegraphics[width=13cm]{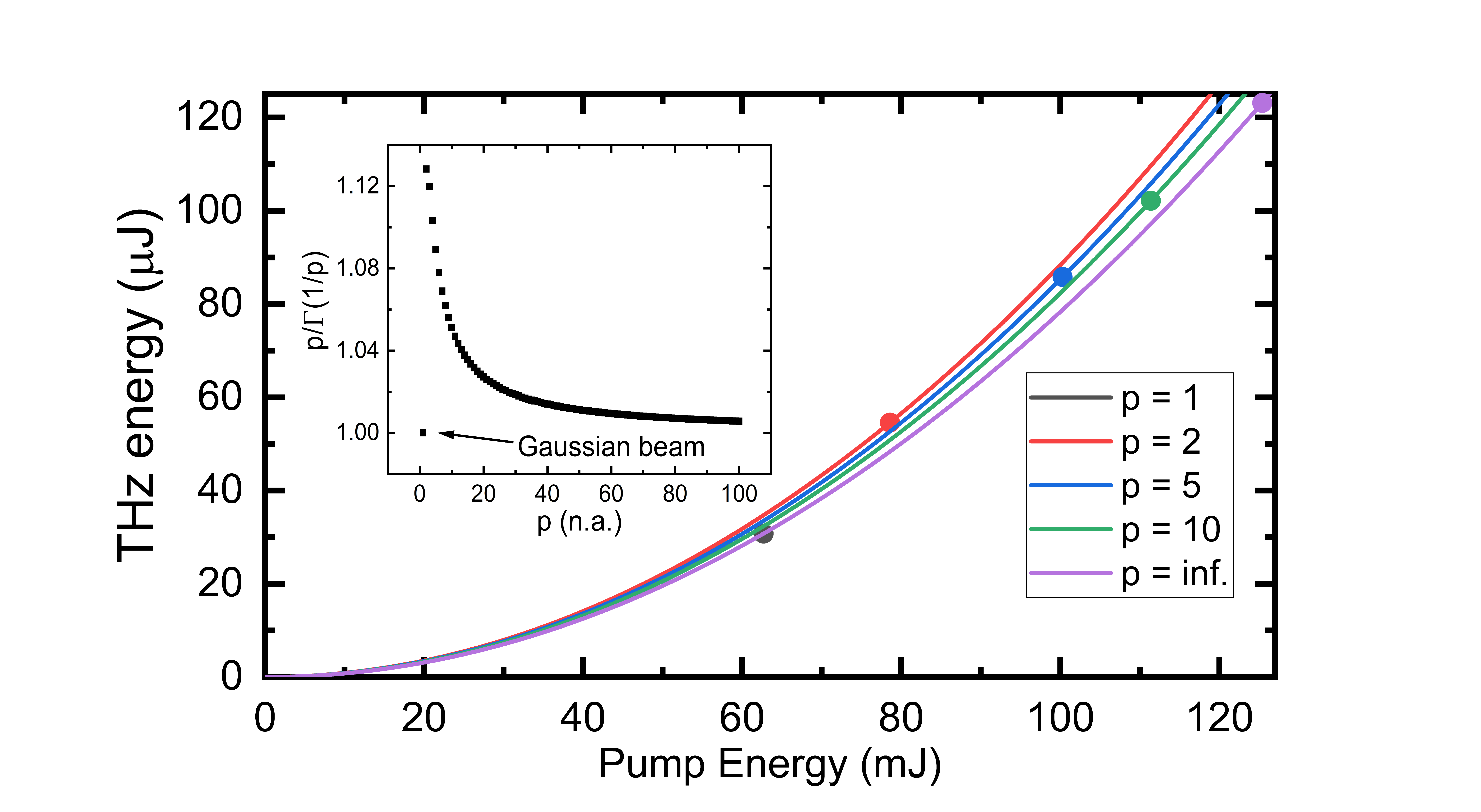}
\caption{Achievable THz pulse energy as a function of the pump energy for different $p-$values assuming 5 mm beam radius and 1.5 ps pump pulse duration. The energy values corresponding to a pump intensity of 100 GW/cm$^2$ are indicated by dots. The difference in the coefficient of the quadratic functions is caused by the $p-$dependence of Eq. 25, which is shown in the inset.}
\end{figure}

\section{Summary}
We have presented a thorough comparison of multicycle THz pulse generation based on lithium niobate in WS and TPFP setups. Pumping WS setups with single pulses, pulse sequences, and intensity-modulated pulses created by the C\&D and the TNB technique were examined, while the TPFP setup was considered as pumped by pulse sequences and intensity-modulated pulses created by the C\&D and the TNB technique. WSs, pumped by FL or chirped single pulses, result in the same temporal shape, spectrum and generation efficiency, but both of these depend on the pump pulse duration. For WSs consisting of five or twenty periods and using 100 GW/cm$^2$ pump intensity, the efficiency has a maximum of close to 0.1\% and 0.3\%, respectively, at 0.15 THz frequency when an optimal pump pulse duration of 1.5 ps is used. On the spectral range of 0.15 – 2.0 THz, for a 5-period WS, the optimal pump duration can be approximated by $\tau_{\mathit{opt}}=\mathrm{\frac{0.245}{\nu_0}}$, where $\nu_0$ is the frequency of the generated THz pulse. The efficiency linearly drops with the THz frequency to 0.05\% at 2.0 THz, because of the strongly increasing absorption.

Pumping by pulse sequence can be used to significantly increase the efficiency. For example, using a pulse sequence consisting of five pulses having the optimal duration found for the single pump pulse case provides about three times larger  efficiency. Efficiencies of almost 1\% can be achieved for a 10-period WS pumped with a 5-pulse sequence of 1.5 ps pulses having an intensity of 100 GW$^2$/cm$^2$. In the case when THz pulses of a specific number of cycles are required, for maximum efficiency, the number of pump pulses should be equal to the number of WS periods. When the number of pump pulses is much higher than the number of wafer stacks, the conversion efficiency is $N_{WS}$ times the single pump pulse case. 

If intensity-modulated pump is used in WS, the TNB and the C\&D techniques give the same results. For TPFP schemes and for short FL pump durations, however, the C\&D technique yields lower efficiencies compared to the TNB technique. Efficiencies obtained with the TPFP are independent of the number of the cycles in the generated THz pulse, while the efficiencies of WSs are basically proportional to the number of cycles (number of wafer pairs). This different behavior, at 0.15 THz for instance, results in an about two times larger efficiency of five-cycle THz pulse generation with TPFP setup than with WS, but the same efficiency for ten-cycle THz pulse generation. Generation of THz pulses having more than ten cycles is more effective using WS than using TPFP setup.  Using TPFP with intensity-modulated pulses allows adjustments of both the THz frequency and the number of cycles by changing the pump pulses, which is not possible for WS sources. 

The presented methods can of course be used at other THz frequencies as well as in other nonlinear crystals. For significantly higher, 2-3 THz frequencies organic crystals or semiconductors can be used. In the case of semiconductors, in order to reduce the multiphoton absorption, it would be beneficial to use longer pump wavelengths. For 10 $\mu$m pump wavelengths, the ZnTe structure period would be 300 $\mu$m, while for GaAs this would be 500 $\mu$m. Thus, for these frequencies, semiconductors can potentially be useful for multicycle THz generation.

\begin{backmatter}
\bmsection{Funding}
Project no. TKP2021-EGA-17 has been implemented with the support provided by the Ministry of Culture and Innovation of Hungary from the National Research, Development and Innovation Fund, financed under the TKP2021 funding scheme.
This work was supported by the National Research, Development and Innovation Office (NKFIH), Hungary, under grant number K147409.
This work was supported by the TWAC project, which is funded by the EIC Pathfinder Open 2021 of the Horizon Europe program under grant agreement 101046504.

\bmsection{Disclosures}
The authors declare no conflicts of interest.

\bmsection{Data availability statement}
Data underlying the results presented in this paper are not publicly available at this time but may be obtained from the authors upon reasonable request.

\bmsection{Supplemental document}
See Supplement 1 for supporting content.

\bibliography{thz}

\end{backmatter}

\end{document}